\documentclass[%
twocolumn,
10pt,tightenlines,
superscriptaddress,
 amsmath,amssymb,
 aps,
prb,
floatfix,
]{revtex4-1}

\usepackage{graphicx}
\usepackage{hyperref}
\usepackage{color}
\usepackage[utf8]{inputenc}
\usepackage{xspace}

\newcommand{\quant}[2]{$#1\,\text{#2}$}
\newcommand{\quantmub}[1]{\quant{#1}{$\mu_\text{Bohr}$/f.u.}}
\newcommand{\michael}[1]{\textbf{\textcolor{blue}{#1}}}
\newcommand{\pascal}[1]{\textbf{\textcolor{red}{#1}}}
\newcommand{\me}[1]{\textbf{\textcolor{green}{#1}}}

\renewcommand{\michael}[1]{}
\renewcommand{\pascal}[1]{}
\renewcommand{\me}[1]{}

\newcommand{\Tc}{T$^\text{c}$\xspace}
\newcommand{\Tp}{T$^\text{p}$\xspace}
\newcommand{\eg}{$e_\text{g}$\xspace}
\newcommand{\eu}{$e_\text{u}$\xspace}
\newcommand{\tgg}{$t_{2\text{g}}$\xspace}
\newcommand{\tu}{$t_{1\text{u}}$\xspace}

\begin{document}

\preprint{}

\title{Ordering tendencies and electronic properties in quaternary Heusler derivatives}

\author{Pascal Neibecker}
\email{pascal.neibecker@frm2.tum.de}
\affiliation{Heinz Maier-Leibnitz Zentrum (MLZ), Technische Universit\"at M\"unchen, 85747 Garching, Germany}
\affiliation{Department of Materials Science, Tohoku University, Sendai 980-8579, Japan}
\author{Markus E. Gruner}
\affiliation{Heinz Maier-Leibnitz Zentrum (MLZ), Technische Universit\"at M\"unchen, 85747 Garching, Germany}
\affiliation{Faculty of Physics and Center for Nanointegration, CENIDE, University of Duisburg-Essen, 47048 Duisburg, Germany}
\author{Xiao Xu}
\author{Ryosuke Kainuma}
\affiliation{Department of Materials Science, Tohoku University, Sendai 980-8579, Japan}
\author{Winfried Petry}
\affiliation{Heinz Maier-Leibnitz Zentrum (MLZ), Technische Universit\"at M\"unchen, 85747 Garching, Germany}
\author{Rossitza Pentcheva}
\affiliation{Faculty of Physics and Center for Nanointegration, CENIDE, University of Duisburg-Essen, 47048 Duisburg, Germany}
\author{Michael Leitner}
\email{michael.leitner@frm2.tum.de}
\affiliation{Heinz Maier-Leibnitz Zentrum (MLZ), Technische Universit\"at M\"unchen, 85747 Garching, Germany}

\date{\today}

\begin{abstract}
The phase stabilities and ordering tendencies in the quaternary full-Heusler alloys NiCoMnAl and NiCoMnGa have been investigated by in-situ neutron diffraction, calorimetry and magnetization measurements. NiCoMnGa was found to adopt the L2$_1$ structure, with distinct Mn and Ga sublattices but a common Ni-Co sublattice. A second-order phase transition to the B2 phase with disorder also between Mn and Ga was observed at \quant{1160}{K}. In contrast, in NiCoMnAl slow cooling or low-temperature annealing treatments are required to induce incipient L2$_1$ ordering, otherwise the system displays only B2 order. Linked to this L2$_1$ ordering, a drastic increase in the magnetic transition temperature was observed in NiCoMnAl, while annealing affected the magnetic behavior of NiCoMnGa only weakly due to the low degree of quenched-in disorder. First principles calculations were employed to study the thermodynamics as well as order-dependent electronic properties of both compounds. It was found that a near half-metallic pseudo-gap emerges in the minority spin channel only for the completely ordered Y structure, which however is energetically unstable compared to the predicted ground state of a tetragonal structure with alternating layers of Ni and Co. The experimental inaccessibility of the totally ordered structures is explained by kinetic limitations due to the low ordering energies.
\end{abstract}

\maketitle
\section{Introduction}\label{sec:Intro}
\subsection{Motivation and Scope}

The class of Heusler alloys, with the ternary system Cu$_2$MnAl as the prototypical representative,\cite{heuslerdpg1903} hosts a variety of systems displaying intriguing properties.\cite{Graf2011} For instance, the latent structural instability in the magnetic Ni$_2$Mn-based compounds gives rise to significant magnetic shape memory\cite{Sozinov2002} and magnetocaloric\cite{Krenke2005} effects. On the other hand, they can also display attractive properties that are directly related to their electronic configuration, with the proposal of spintronics, which relies on the detection and manipulation of spin currents, as an example. In a magnetic tunnel junction for instance, the achievable tunneling magnetoresistive effect and thereby the miniaturization of components depends on the spin polarization of the conduction electrons in the electrodes\cite{Julliere1975}. As a consequence, half-metallic materials, which have a 100\% spin polarization due to a band gap at the Fermi level in one spin channel, are highly sought after and currently the focus of both theoretical and experimental investigations.

While the first half-metal identified by theoretical calculations in 1983 by Groot et al.\cite{Groot1983} was the half-Heusler compound NiMnSb of C1$_\text{b}$ structure, also in full-Heusler alloys with L2$_1$ structure half-metallic properties have been predicted\cite{Ishida1982,Fujii1990} and experimentally observed\cite{Jourdan2014}. Recently, also a large number of quaternary Heusler derivatives, among them NiCoMnAl\cite{Halder2015} and NiCoMnGa\cite{Alijani2011}, have been suggested by \textit{ab initio} calculations to be half-metals in their fully ordered Y structure \cite{Ozdougan2013}. Half-metallic properties in the Ni$_{2-x}$Co$_x$MnAl\cite{Okubo2011} and  Ni$_{2-x}$Co$_x$MnGa\cite{Kanomata2009} systems have additionally been proposed for the Co-rich side of the respective phase diagrams on the basis of magnetization measurements via the Generalized Slater-Pauling rule. It is obvious that the degree of chemical order will have direct consequences for the half-metallic properties of these systems. However, the connection between atomic order, segregation and functional properties has also been established for the magnetocaloric and metamagnetic shape memory effects,\cite{Recarte2012,Barandiaran2013b}, ferroic glasses\cite{Monroe2015} and the recently reported shell-ferromagnetism in off-stoichiometric Heusler compounds.\cite{Cakir2016}

\begin{figure*}[t]\centering
\includegraphics{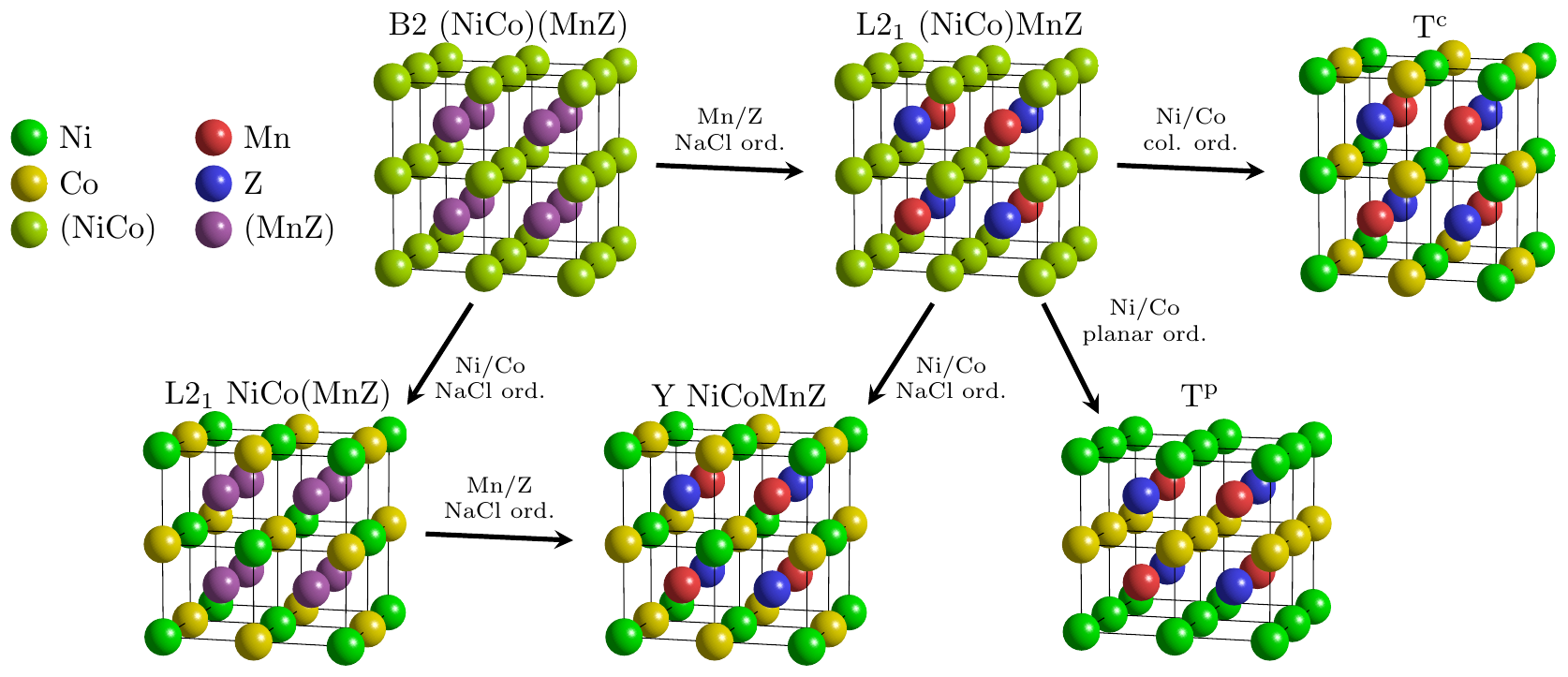}
\caption{Illustration of the ordered structures of NiCoMnZ considered here. Starting with B2 (NiCo)(MnZ), NaCl-type ordering on either sublattice leads to L2$_1$ structure of (NiCo)MnZ or NiCo(MnZ) type, and ordering on both sublattices to the Y structure. Those four structures have cubic symmetry. Ordering of the Ni and Co atoms in L2$_1$ (NiCo)MnZ into alternating columns or planes gives the tetragonal \Tc and \Tp structures, respectively.}
\label{fig:structures}
\end{figure*}

In assessing the potential of a given material for application following from its electronic structure, theoretical and experimental investigations have contrasting characteristics: in \textit{ab initio} calculations, the distribution of electronic charge is the fundamental quantity that is considered, which depends in principle only on the positions of the ions and their atomic numbers. From this, other important properties can be derived, like total energies, magnetic moments and forces on the ions. Different structures can be compared in terms of their total energies, but chemical disorder must be taken into account appropriately. This can be handled very efficiently in terms of the coherent potential approximation (see Ref.{} \cite{Ruban2008} for a recent review), which, however, does not provide an easy way to account for ionic relaxations. On the other hand, explicit calculations of disordered structures with randomly distributed atoms in larger super-cells are much more involved and numerically intensive. Thus, for practical reasons often an ordered configuration is assumed to be representative. On the other hand, in experiments the state of order in the sample is relevant for the potential application, while the determination of aspects of the electronic structure is often quite hard, which especially applies for the spin polarization. Thus, it seems indicated to combine the respective strengths of experiment and theory, which is what we set out to do in this paper. Specifically, in the systems of NiCoMnAl and NiCoMnGa we study the degrees of equilibrium long-range order and the associated order/disorder phase transitions by \textit{in situ} neutron diffraction, and the kinetics of order relaxation during isothermal annealing by way of its effect on magnetization and Curie temperature. Further, we perform \textit{ab initio} calculations on different ordered and disordered structures to determine the associated electronic structures as well as ordering energies. As we will show, these calculations imply that among the realistic candidates only the hitherto assumed Y ordering displays half-metallicity, but does not correspond to the actual ground state. In addition, the associated ordering energies are small, which explains the experimentally observed stability of disorder among Ni and Co. 

\subsection{States of order in quaternary Heusler derivatives}\label{subsect:structures}
To facilitate the discussion of the different ordered quaternary structures and their relations later in this article, we enumerate here the structures, define the nomenclature and summarize the pertinent knowledge on their ternary parent compounds. 

Heusler alloys in the strict sense of the word are ternary systems of composition X$_2$YZ displaying L2$_1$ order, which is defined by the space group 225 (Fm$\bar{3}$m) with inequivalent occupations of the Wyckoff positions 4a, 4b and 8c. Typically, X is a late transition metal occupying preferentially 8c, while an early transition metal Y and a main-group metal Z occupy the other two sites,\cite{Graf2011} with Cu$_2$MnAl the prototypical representative. 

Ni$_2$MnGa conforms to above definition and displays a stable L2$_1$ phase at intermediate temperatures.\footnote{We neglect here the martensitic transitions below room temperature.} Around \quant{1053}{K} it shows a second-order disordering transition to the B2 (CsCl) structure,\cite{SanchezAlarcos2007} corresponding to a mixing of Mn and Ga, that is, it acquires space group 221 (Pm$\bar{3}$m) with Wyckoff position 1a occupied preferentially by Ni, while Mn and Ga share position 1b. This partial disordering can be understood by the observation that B2 order, i.e., the distinction between Ni on the one hand and Mn and Ga on the other hand, is stabilized by nearest-neighbor interactions on the common bcc lattice, while the ordering between Mn and Ga corresponding to full L2$_1$ order can only be effected by the presumably weaker next-nearest-neighbor interactions. Indeed, in Ni$_2$MnAl only the B2 state or at the most very weak L2$_1$ order can experimentally be observed.\cite{Acet2002} In both systems the B2 state is stable up to the melting point, that is, there is no transition to the fully disordered bcc state. In the Co-based systems, the situation is remarkably similar, with well-developed L2$_1$ order in Co$_2$MnGa and only B2 order in Co$_2$MnAl.\cite{websterjphyschemsolids1971}

It seems probable, and is indeed corroborated by our experimental observations to be reported below, that the behavior of the quaternary systems NiCoMnGa and NiCoMnAl can be traced back to their ternary parent compounds. The most plausible candidates of ordered structures following this reasoning are illustrated Fig.~\ref{fig:structures}. Given that both Ni- and Co-based ternary parents display the B2 structure at high temperatures, it is natural to assume this to be also the case for NiCoMnZ, with site 1a shared by Ni and Co and site 1b by Mn and Z. We will denote this as (NiCo)(MnZ), where the parentheses denote mixing between the enclosed elements.\footnote{We do not consider the other two B2 possibilities (NiMn)(CoZ) and (NiZ)(CoMn).}

As temperature is decreased, transitions to states of higher order can appear. For the Mn-Z sublattice, a NaCl-like ordering of Mn and Z is most likely by analogy with the ternary parents. Assuming the same kind of interaction favoring unlike pairs also between Ni and Co, the realized structures depend on the relative strengths: for dominating Mn-Z interactions, the B2 phase would transform to an L2$_1$ structure of type (NiCo)MnZ, where Ni and Co are randomly arranged over the 8c sites, and in the converse case to L2$_1$ NiCo(MnZ) with Mn and Z on 8c. In either case, the ordering of the other sublattice at some lower temperature would transform the system to the so-called Y structure\cite{Pauly1968,Bacon1971} of prototype LiMgPdSn\cite{Eberz1980} with space group 216 (F$\bar{4}$3m) and Wyckoff positions 4a, 4b, 4c, and 4d being occupied by Ni, Co, Mn and Z, respectively. 

However, as the kind of chemical interaction within the Ni-Co sublattice is as yet unknown, also other possibilities have to be considered. In principle, there is an unlimited number of superstructures on the L2$_1$ (NiCo)MnZ structure, corresponding to different Ni/Co orderings. In particular, apart from the above-mentioned cubic Y structure (with NaCl-type Ni/Co ordering) there are two other structures with a four-atom primitive cell, making them appear \textit{a priori} equally likely to be realized as the Y structure. These are tetragonal structures characterized by either alternating columns or planes of Ni and Co atoms, which we denote by \Tc and \Tp. Specifically, the \Tc structure has space group 131 (P4$_2$/mmc), with Ni on Wyckoff position 2e, Co on 2f, Mn 2c, and Z on 2d, while \Tp has space group 129 (P4/nmm) with Ni on 2a and Co on 2b, while Mn and Z reside on two inequivalent 2c positions, with prototype ZrCuSiAs\cite{Johnson1974}. Note that the Ni/Co ordering in these three fully-ordered structures can equally be understood as alternating planes in different crystallographic orientations, with \Tp corresponding to $(1,0,0)$ planes, \Tc to $(1,1,0)$, and Y to $(1,1,1)$ planes. Finally, of course the possibility of phase separation into L2$_1$ Ni$_2$MnZ and Co$_2$MnZ has to be considered.

\section{Macroscopic properties}\label{sec:mag}
\subsection{Sample preparation and thermal treatments}
Nominally stoichiometric NiCoMnAl and NiCoMnGa alloys have been prepared by induction melting and tilt casting of high-purity elements under argon atmosphere. After casting, the samples have been subjected to a solution-annealing treatment at \quant{1273}{K} followed by quenching in room-temperature water. In this state, the samples have been checked for their actual composition using wavelength-dispersive X-ray spectroscopy (WDS). For each alloy, eight independent positions have been measured. The average over the retrieved values are given in Table \ref{wds}, showing satisfactory agreement with the nominal compositions. Additionally, sample homogeneity was confirmed by microstructure observation using backscattered electrons. 

\begin{table}[b]
\centering
\begin{tabular}{r||c|c|c|c|c}
at. $\%$($\pm 0.5 \%$)  & Ni & Co & Mn & Ga & Al\\
\hline
 NiCoMnGa& 25.6  & 23.4 & 26.1 & 25.0&- \\
 NiCoMnAl& 25.1 & 25.5 & 25.7 & - & 23.8
\end{tabular}
\caption{Composition as determined by WDS.}
\label{wds}
\end{table}

In order to track the ordering processes of the alloys upon low-temperatures isothermal aging, samples have been annealed at \quant{623}{K} for different times and water-quenched. Thus, for both systems we consider four states, corresponding to the as-quenched state and after annealings for \quant{6}{h}, \quant{24}{h}, and \quant{72}{h}, respectively, denoted in the following by \textsc{aq}, \textsc{ann6h}, \textsc{ann24h}, and \textsc{ann72h}. Previous results \cite{Neibecker2014} have proven this low-temperature annealing protocol to be successful for increasing the achievable state of order in structurally similar alloys of the Ni$_2$MnAl system.  

\subsection{Magnetization measurements}
Magnetization measurements corresponding to the different annealing conditions were performed, specifically the Curie temperatures $T_\text{C}$ and spontaneous magnetization values $M_\text{S}$ have been determined. Temperature-dependent magnetization measurements were done in a TOEI Vibrating Sample Magnetometer (VSM) applying an external magnetic field of \quant{500}{Oe} in a temperature range from room temperature to \quant{693}{K}. The spontaneous magnetization for NiCoMnGa has been determined with a Superconducting Quantum Interference Device (SQUID) based Quantum Design MPMS system at \quant{6}{K} employing external magnetic fields up to \quant{7}{T}. Since for the ductile NiCoMnAl alloy sample preparation turned out to have an effect on sample properties, presumably due to introduced mechanical stresses, in this alloy system temperature-dependent magnetization measurements have been performed by VSM on samples of larger size in an external field of \quant{1.5}{T}.

\begin{figure}[t]\centering
\includegraphics{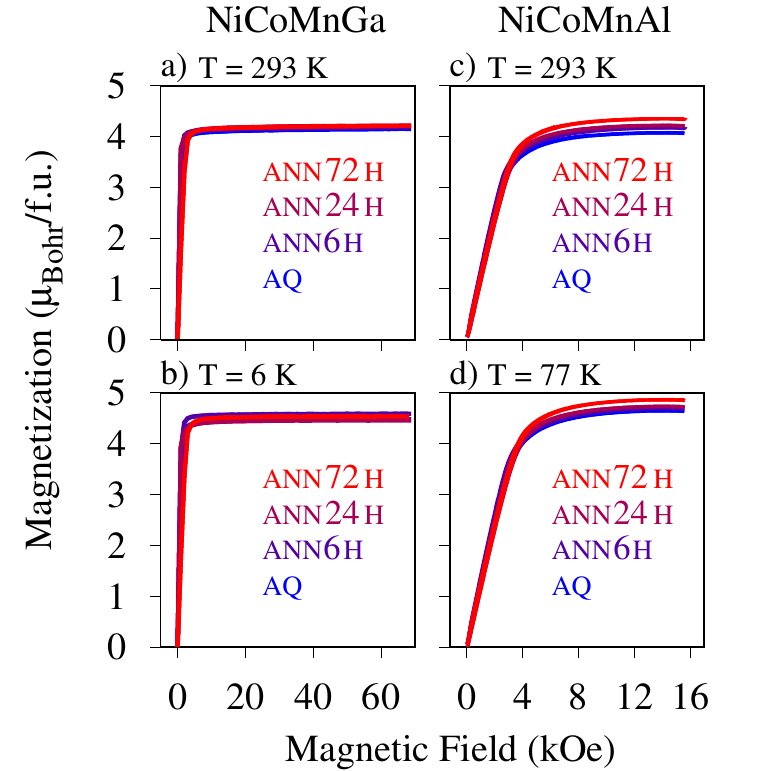}
\caption{Field dependent magnetization of NiCoMnGa in different annealing conditions measured at \quant{293}{K} a) and \quant{6}{K} b), as well as for NiCoMnAl measured at \quant{293}{K} c) and \quant{77}{K} d).}
\label{fig:Mag}
\end{figure}

\begin{table*}[tb]
\centering
\begin{tabular}{c|c|cc|ccc}
 && \multicolumn{2}{c|}{$T_\text{C}$ (K) [VSM]} & \multicolumn{3}{c}{$M_\text{S}$ ($\mu_{\text{Bohr}}$/f.u.)}  \\
  &state& up & down & \quant{6}{K} [SQUID] &  & \quant{293}{K} [SQUID] \\

\hline
 NiCoMnGa &\textsc{aq}& 636.0 & 633.9 & 4.47 &   &4.11   \\
  &\textsc{ann6h}& 635.4 & 632.3   & 4.58 &  & 4.14  \\
  &\textsc{ann24h}& 637.1 & 633.4   & 4.46 &  & 4.18 \\
  &\textsc{ann72h}& 638.2 & 632.9   & 4.53&  & 4.17 \\
  \hline
  \hline
  &  & up & down & \quant{0}{K} [VSM extrap.] & \quant{77}{K} [VSM] & \quant{293}{K} [VSM] \\
  \hline
  NiCoMnAl &\textsc{aq}& 572.1 & 583.3  & 4.68  & 4.61 & 3.98  \\
  &\textsc{ann6h}& 586.8 & 590.2  &  4.72 & 4.66 & 4.08 \\
  &\textsc{ann24h}& 593.2 & 593.8 &  4.75 & 4.72 & 4.15 \\
  &\textsc{ann72h}& 602.9 & 599.2 &  4.89 & 4.86 & 4.30 \\
\end{tabular}
\caption{Magnetic properties of NiCoMnAl and NiCoMnGa in different annealing conditions measured by VSM and SQUID.}
\label{tab2}
\end{table*}

\begin{figure*}\centering
\includegraphics{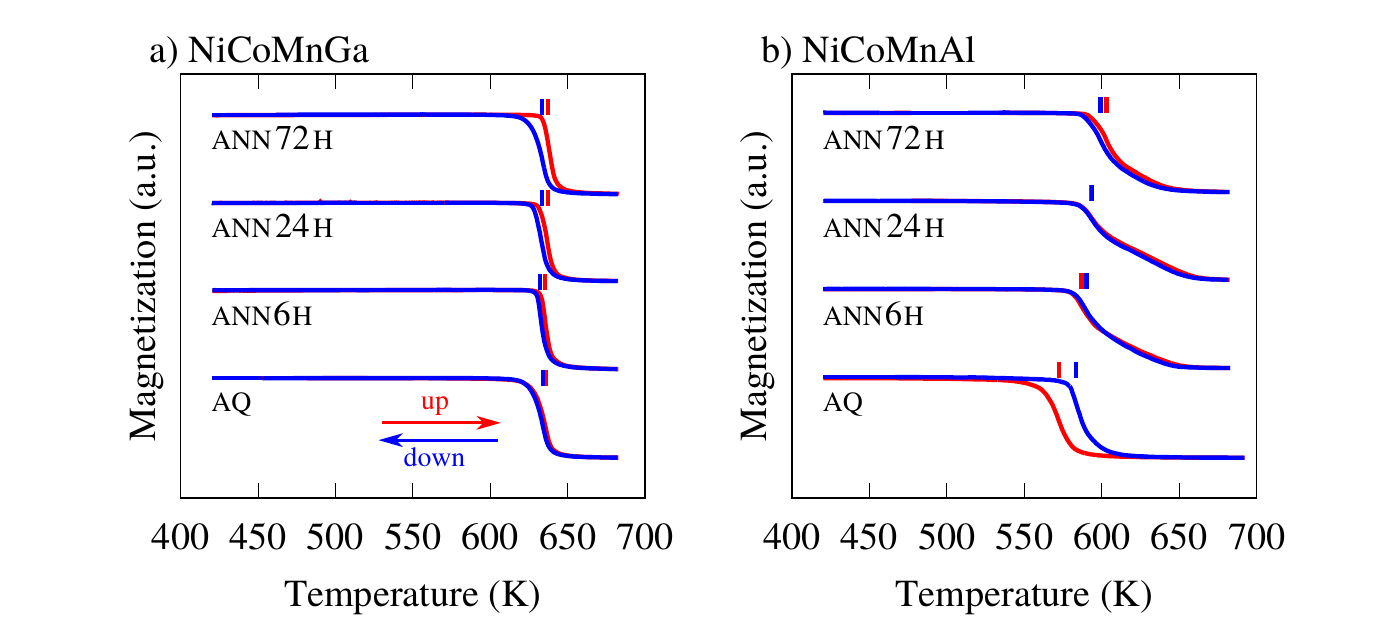}
\caption{Temperature-dependent magnetization of NiCoMnGa a) and NiCoMnAl b) in different annealing conditions at an external magnetic field of \quant{500}{Oe}. Depicted in red and blue are the heating and cooling curves, while the markers represent the retrieved values for the Curie temperatures $T_\text{C}$.}
\label{fig:MT}
\end{figure*}

Figure \ref{fig:Mag} a) and b) shows field-dependent magnetization curves ($M(H)$) of NiCoMnGa in four different annealing conditions measured at \quant{293}{K} and \quant{6}{K}. The spontaneous magnetization $M_\text{S}$ has been retrieved via constructing Arrott plots. The obtained values are given in Table \ref{tab2}. While, as expected, the spontaneous magnetization increases with decreasing measurement temperature, no apparent effects on $M_\text{S}$ due to annealing are visible, with the values of $M_\text{S}$ scattering around $4.15$ and \quantmub{4.50} at \quant{293}{K} and \quant{6}{K}, respectively. These values show reasonable agreement to previous studies with $M_\text{S}$ being stated as \quantmub{4.67} at a temperature of \quant{5}{K}.\cite{Kanomata2009} 

Figure \ref{fig:Mag} c) and d) show $M(H)$ curves of NiCoMnAl in four different annealing conditions measured at \quant{293}{K} and \quant{77}{K}. Additionally, $M(T)$ curves of NiCoMnAl in the four annealing conditions at an external magnetic field of \quant{15}{kOe} that were used to extrapolate the \quant{0}{K} value of the spontaneous magnetization $M_\text{S}$ are given in the Supporting Information. The determined values for $M_\text{S}$ are listed in Table \ref{tab2} and range from \quantmub{4.68} in the \textsc{aq} state to \quantmub{4.89} in the \textsc{ann72h} state. This increase of \quantmub{0.22} during annealing in NiCoMnAl is significantly larger than the corresponding effect in NiCoMnGa. Clearly, the difference in $M_\text{S}$ with annealing further increases at higher measurement temperatures due to lower magnetic transition temperatures in the shorter annealed samples. The values obtained for $M_\text{S}$ are in good agreement with two previous studies where $M_\text{S}$ in the B2 ordered state was reported as \quantmub{4.66}\cite{Halder2015} and \quantmub{4.90}\cite{Okubo2011}

Figure \ref{fig:MT} shows the corresponding temperature-dependent magnetization measurements ($M(T)$) of NiCoMnGa and NiCoMnAl under an external field of \quant{500}{Oe}. The magnetization curves are normalized since absolute magnetization values are, due to sample shape-specific demagnetization fields, not meaningful under low external magnetic fields. In the following discussion, we define the apparent Curie temperature as the locus of the maximal slope of the $M(T)$ curves.

In NiCoMnAl, the magnetic transition temperature increases from \quant{572.1}{K} to \quant{602.9}{K} with annealing of the samples, reflecting a corresponding increase of L$2_1$ order. The specific transition temperatures are given in Table \ref{tab2}. This compares satisfactorily with the value of \quant{570}{K} quoted by Okubo et al.\cite{Okubo2011} for samples quenched from the B2 region. An important point to note is that, in order to probe the high Curie temperatures in these systems, during the measurements the sample is subjected to temperatures where the ordering kinetics become appreciable. Specifically, with ordering kinetics at \quant{623}{K} on the order of hours, the Curie temperatures below \quant{600}{K} measured on heating at a rate of \quant{2}{K/min} can safely be assumed to correspond to the degree of order imposed by the isothermal annealing treatments. However, at the maximum temperature of \quant{683}{K} the degree of order will relax during the measurement towards the corresponding equilibrium value, leading to an increase of order for the \textsc{aq} sample and a decrease when starting from a high degree of order. This difference between heating and cooling curves is well discernible.

NiCoMnGa shows a magnetic transition from the ferromagnetic to the paramagnetic state between 636.0 and \quant{638.2}{K}. Since the degree of L$2_1$ order in NiCoMnGa is high in all annealing conditions, annealing has a much smaller effect on $T_\text{C}$ than in NiCoMnAl. The determined transition temperatures are given in Table \ref{tab2}. Apparently, still a small increase of order with annealing exists in this alloy system. We interpret the constant offset of about \quant{3.5}{K} between heating and cooling to effects of thermal inertia.

\subsection{Differential scanning calorimetry}
\begin{figure}[tb]\centering
\includegraphics{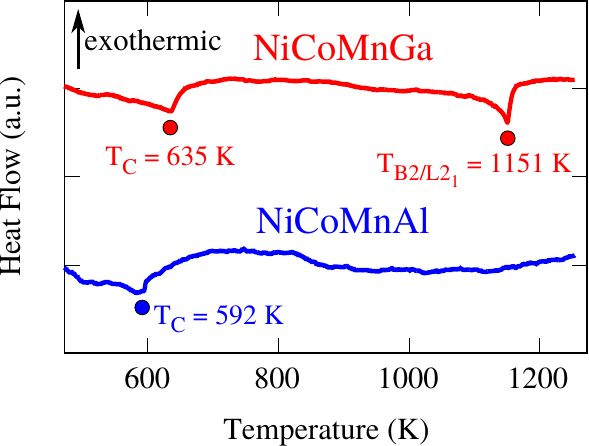}
\caption{DSC measurements of annealed NiCoMnGa and NiCoMnAl. DSC curves have been recorded on heating with a rate of \quant{10}{K/min}.}
\label{fig:DSC}
\end{figure}

Differential Scanning Calorimetry (DSC) has been employed to analyze both alloys with respect to magnetic and structural phase transitions on a Netzsch DSC 404 C Pegasus. All measurements have been performed at a heating rate of \quant{10}{K/min} over a temperature range from 300 K to 1273 K. Figure \ref{fig:DSC} shows DSC results for the NiCoMnGa and NiCoMnAl alloys that have been subject to solution annealing at \quant{1273}{K}, quenching to room temperature, followed by a low-temperature annealing  at \quant{623}{K} applied with the intention to adjust a large degree of L2$_1$ order. Taking into account different ordering kinetics, the NiCoMnGa alloy was annealed for \quant{24}{h}{K}, while the NiCoMnAl alloy was annealed for \quant{72}{h}. 

Both alloys show a clear magnetic transition from the ferromagnetic to the paramagnetic state at \quant{592}{K} and \quant{635}{K} for NiCoMnAl and NiCoMnGa, respectively. Those values show good agreement to the values obtained by magnetization measurements (Table \ref{tab2}), justifying our approach of defining the Curie temperatures via the position of maximal slope in the magnetization under constant field. NiCoMnGa shows additionally an order-disorder phase transition at higher temperatures that can be assigned according to our neutron diffraction measurements (Sec. \ref{sec:neutron}) to the transition from the L2$_1$-(NiCo)MnGa to the B2-(NiCo)(MnGa) structure, which is in accordance with the behavior of the structurally similar Ni$_2$MnGa compound\cite{SanchezAlarcos2007} and with previous results from Kanomata et al.\cite{Kanomata2009} The phase transition temperature was determined as \quant{1151}{K}, a value in excellent agreement to the \quant{1152}{K} reported in Ref.{} \onlinecite{Kanomata2009}. In contrast, NiCoMnAl does not show any further apparent peaks in the calorimetric signal besides the magnetic transition. 

\begin{figure*}[tb]\centering
\includegraphics{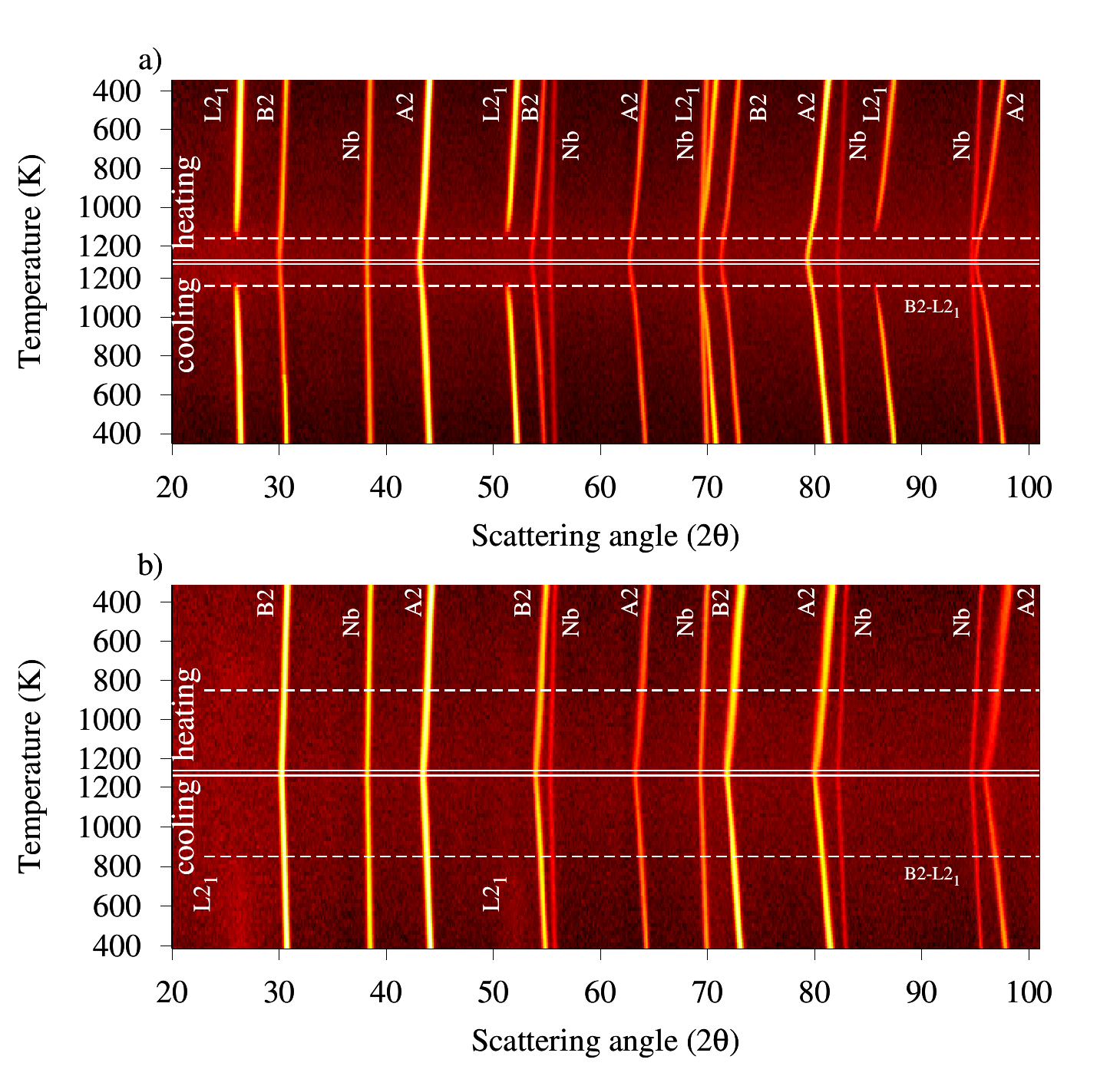}
\caption{Waterfall plots of the temperature-dependent neutron powder diffraction patterns of NiCoMnGa (above) and NiCoMnAl (below) on heating and cooling with rates of approximately \quant{2}{K/min}. Before the measurement, samples were solution annealed at \quant{1273}{K} and quenched in room temperature water.}
\label{fig:NiCoMnGa_water}
\end{figure*}

\section{Neutron diffraction}\label{sec:neutron}
Neutron diffraction measurements have been performed at the SPODI\cite{SPODI2015} high-resolution neutron powder diffractometer at the Heinz Maier-Leibnitz Zentrum (MLZ) in Garching, Germany. Polycrystalline samples were measured continuously on heating and cooling between room temperature and \quant{1273}{K}, employing rates of approximately \quant{2}{K/min} and a recording frequency of approximately one pattern per 15 minutes. Measurements have been done using Nb sample holders and employing a neutron wavelength of \quant{1.5406}{\AA}. Temperature-dependent lattice constants, peak widths and structure factors corresponding to the different degrees of long-range order have been refined. Additionally, for the depiction of the waterfall plots, data treatment as described in Ref.{} \onlinecite{Hoelzel2012} has been applied. 

Figure \ref{fig:NiCoMnGa_water} shows waterfall plots of the neutron diffraction patterns of NiCoMnGa/Al upon heating and cooling on a logarithmic pseudocolor scale. All reflection families, namely L2$_1$, B2 and A2, as well as the peaks due to the Nb sample holder, are labeled in the figure. Their presence correspond to the symmetry breaking into inequivalent sublattices as discussed in Sec. \ref{sec:Intro}, and their strength indicates the quantitative degree of long-range order. The A2 peaks are not influenced by any disorder in the system, since here all lattice sites contribute in phase. The presence of the B2 peak family indicates different average scattering lengths on the Ni-Co and the Mn-Z sublattices. Finally, L2$_1$ peaks are due to a further symmetry breaking between either the 4a and 4b and/or 4c and 4d sublattices. Note that such a qualitative reasoning cannot distinguish whether the system has the Y structure or one of the two possible L2$_1$ structures, which can only be decided by a quantitative analysis (as will be done below). Similar as for the A2 peak family, the intensity of the B2 peak family is not influenced by the degree of L2$_1$ order.

In the waterfall depiction, the evolution of peak position (in qualitative terms) peak intensity with temperature can be followed nicely. Initially, the samples correspond to the state quenched from \quant{1273}{K}. Already in this state, NiCoMnGa exhibits L2$_1$ order as evidenced by the presence of the corresponding diffraction peaks. Upon heating, first of all the thermal expansion of the lattice is observed with the peak positions shifting to smaller scattering angles. Simultaneously, the peaks stemming from the Nb sample holder can clearly be distinguished from the sample peaks due to their lower rate of thermal expansion. At approximately \quant{1160}{K}, a disordering phase transition from the L2$_1$ phase to the B2 phase is observed. This is reversed on cooling at nearly the same temperature, which shows that at these high temperatures the equilibrium states of order are followed closely. The observed value of \quant{1160}{K} is in good agreement to the \quant{1151}{K} determined by calorimetry.

\begin{figure}[t]\centering
\includegraphics{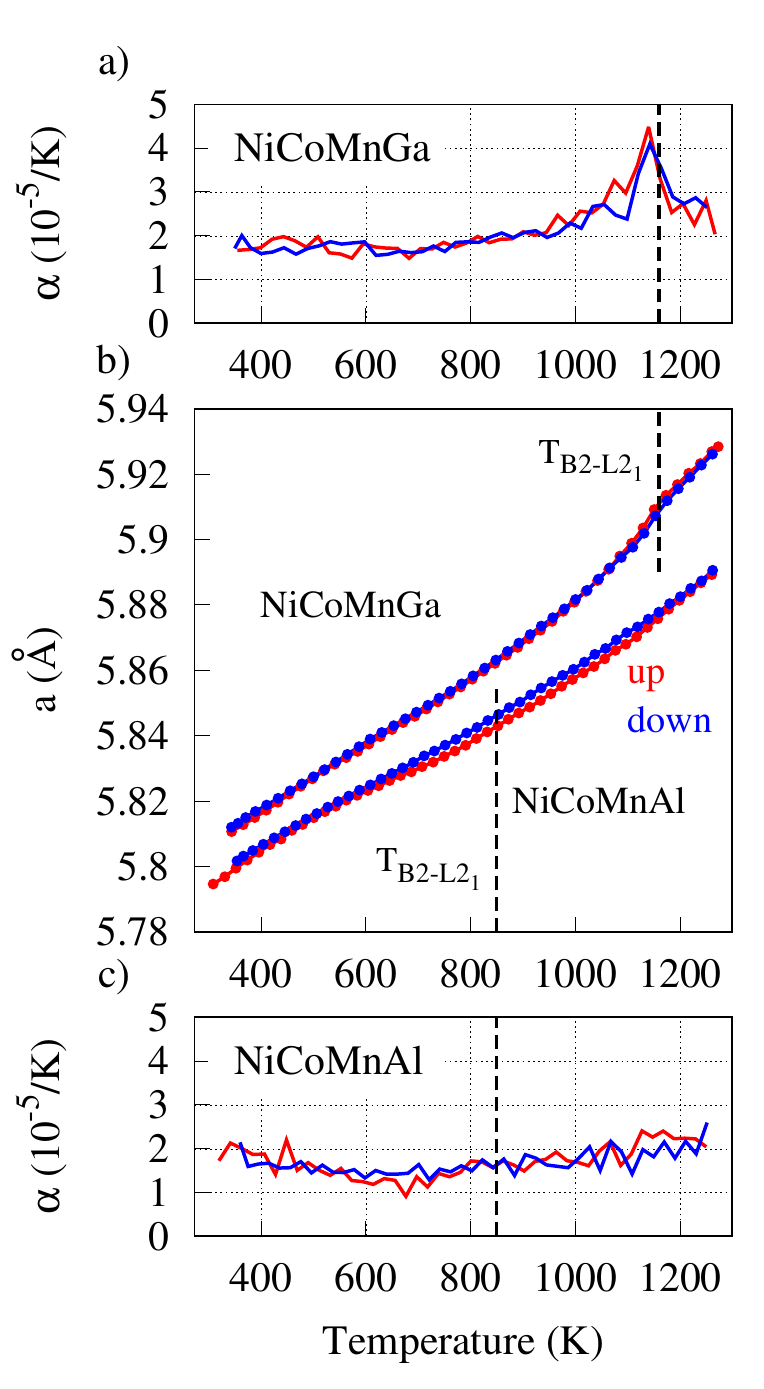}
\caption{b) Temperature-dependent lattice constants of NiCoMnGa and NiCoMnAl on heating and cooling starting with samples quenched from \quant{1273}{K}. Upper and lower panels depict the temperature-dependent thermal expansion coefficients $\alpha = \frac{1}{a}\frac{\mathrm{d}a}{\mathrm{d}T}$ for NiCoMnGa a) and NiCoMnAl c), respectively.}
\label{fig:lattice_constants} 
\end{figure}

In contrast to NiCoMnGa, NiCoMnAl is found to have a B2 state in the as-quenched condition with no L2$_1$ reflections visible. On cooling, the peaks are slightly narrower than on heating, indicating the release of internal stresses in the sample remaining from quenching. Interestingly, upon slow cooling the sample down from \quant{1273}{K}, at approximately \quant{850}{K} very diffuse maxima are appearing at the positions where L2$_1$ reflections would be expected. Numerical analysis of the corresponding regions on heating suggests that also already here a very weak intensity is found as soon as temperature regions are reached that are sufficient to facilitate a relaxation of order via diffusion. \michael{das muss ich definitiv noch machen, weil nach hinten das ja als wohldefinierter übergang dargestellt wird, sonst würde das zu beliebig aussehen.} Arguably, the diffuse intensity observed is the manifestation of L2$_1$ short-range order or incipient L2$_1$ long-range order with very small anti-phase domains. Such anti-phase domains have previously been observed in Ni$_2$MnAl$_{0.5}$Ga$_{0.5}$ alloys,\cite{Ishikawa2008,Umetsu2011} where the phase transition temperature implies ordering kinetics on experimentally accessible time scales. The absence of well-defined L2$_1$ order as well as the pronouncedly lower B2-L2$_1$ transition temperature in NiCoMnAl compared to the NiCoMnGa alloy is consistent with the behavior observed in the related Ni$_2$MnAl and Ni$_2$MnGa compounds where transition temperatures of, respectively, \quant{775}{K}\cite{Kainuma2000} and \quant{1053}{K}\cite{SanchezAlarcos2007} have been reported.

While confirming a state of B2 order, in neutron diffraction no magnetic superstructure peaks are observed. Thus, in contrast to Ni$_2$MnAl, where Ziebeck et al. \cite{Ziebeck1975} discovered a helical magnetic structure manifesting itself in form of antiferromagnetic superstructure reflections and satellite peaks at the (200) and (220) reflections, NiCoMnAl is entirely ferromagnetic even under B2 order.  This goes along with $M(H)$ measurements (Sec. \ref{sec:mag}) showing prototypical ferromagnetic properties. In contrast, for Ni$_2$MnAl, antiferromagnetic properties haven been reported.\cite{Acet2002} Presumably, the drastic difference in magnetic structure results from strong ferromagnetic interactions in the system introduced by Co, overcoming the antiparallel coupling between neighboring Mn atoms.

Figure \ref{fig:lattice_constants} shows the temperature-dependent lattice constants retrieved from fitting the in-situ neutron diffraction data as well as the corresponding temperature-dependent thermal expansion coefficients. At \quant{340}{K}, the lattice constant of approximately \quant{5.81}{\AA} in as quenched NiCoMnGa is only slightly larger than the one of NiCoMnAl with approximately \quant{5.80}{\AA}, while the thermal expansion is similar in both alloys with a value of approximately \quant{2\times 10^{-5}}{K$^{-1}$}. In the case of NiCoMnGa, the heating and cooling curves coincide, indicating little effect of the applied quenching treatment. The B2-L2$_1$ transition is clearly mirrored in the lattice constant, with a maximum in the thermal expansion coefficient around \quant{1160}{K}, in agreement with the calorimetric transition temperature and the vanishing of L2$_1$ intensities in neutron diffraction. 

\begin{figure*}[t]\centering
\includegraphics{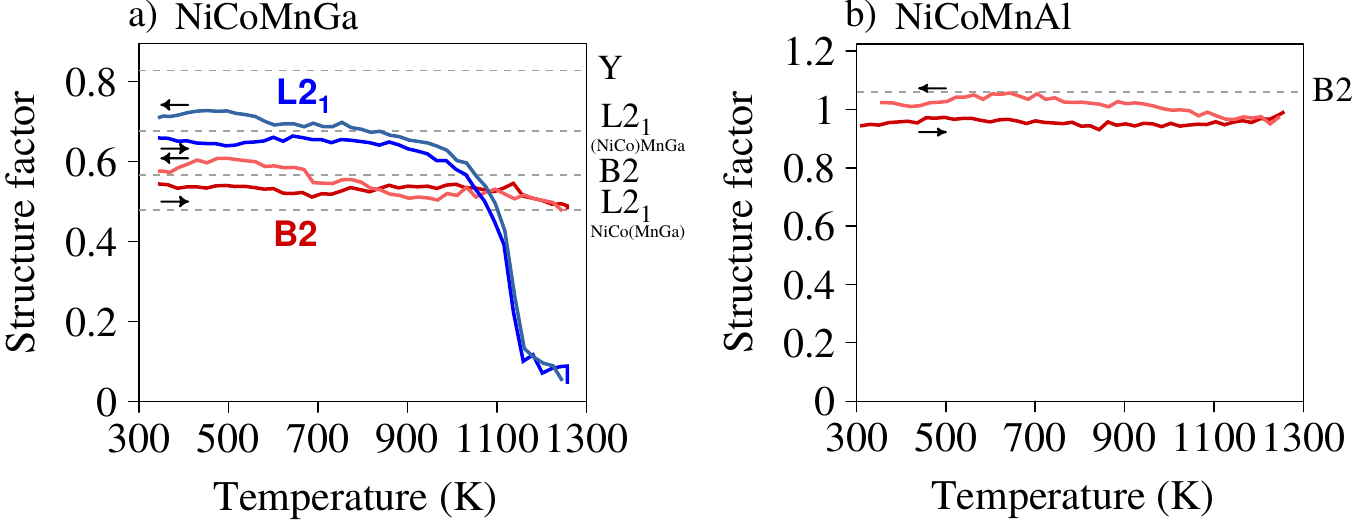}
\caption{Structure factors of the B2 and L2$_1$ peak families as function of temperature as determined by in-situ neutron diffraction in NiCoMnGa a) and NiCoMnAl b).}
\label{fig:struc_facts}
\end{figure*}

In the case of NiCoMnAl, the B2--L2$_1$ ordering transition is neither visible directly in the lattice constant nor in the thermal expansion coefficient. However, this system displays another striking effect with the divergence of the lattice constants on heating and cooling at intermediate temperatures. The absence of an analogous effect in the determined Nb lattice constants proves that this deviation is real as opposed to, e.g., an error in the determination of the sample temperature. We interpret it to be due to a superposition of a lattice expansion due to quenched-in disorder with a lattice contraction due to a quenched-in vacancy supersaturation. On heating, around \quant{650}{K} ordering kinetics become active, leading to a relaxation of the lattice expansion, while only at temperatures above \quant{1100}{K} vacancies become mobile enough to equilibrate their concentrations at vacancy sinks such as surfaces or grain boundaries. Thus, in this interpretation the agreement in the lattice constants of the slow-cooled and quenched states at low temperatures is just a coincidence.

Figures \ref{fig:struc_facts} show the temperature-dependent structure factors of NiCoMnGa and NiCoMnAl, i.e., essentially the ratio of the intensities of the B2 and L2$_1$ peaks to the A2 peak families after taking into account Lorentz factors and Debye-Waller factors. The theoretical structure factors for different kinds of disorder are depicted in the figures as stroked lines. In the case of NiCoMnGa, the second-order B2--L2$_1$ transition at \quant{1160}{K} is clearly visible. Additionally, this evaluation gives credence to the scenario of the observed L2$_1$ intensity being due to solely Mn/Ga order as opposed to Y ordering or only Ni/Co ordering. Interestingly, with increasing temperature also the degree of B2 order decreases somewhat in both systems. Also, the degrees of order on cooling are always higher than on heating, which is indicative of some amount of disorder after quenching. These qualitative conclusions seem valid even though quantitative interpretations of the data have to be treated with caution considering the limited number of crystallite grains fulfilling the Bragg condition that defines the statistical precision.

\section{First-principles calculations}\label{sec:theory}
We performed \textit{ab initio} calculations for the structures proposed in Sect.~\ref{subsect:structures}. Specifically, we computed ordering energies and electronic densities of states (DOS) by plane-wave density functional theory as implemented in VASP (Vienna Ab-initio Simulation Package),\cite{VASP1} and magnetic interactions in the Liechtenstein approach\cite{Liechtenstein1987} as implemented by Ebert et al. \cite{Ebert2011} in their Korringa-Kohn-Rostoker Green's function code (SPR-KKR). 

\subsubsection{Computational details}
In the VASP calculations, the disordered structures were realized by 432 atom supercells (corresponding to 6$\times$6$\times$6 bcc cells) with random occupations, taking advantage of the efficient parallelization in VASP for massively parallel computer hardware. Here, the wavefunctions of the valence electrons are described by a plane wave basis set, with the projector augmented wave approach taking care of the interaction with the core electrons.\cite{VASP2} Exchange and correlation was treated in the generalized gradient approximation using the formulation of Perdew, Burke and Ernzerhof.\cite{Perdew1996} We converged unit cell dimensions and atomic positions by a conjugate gradient scheme until forces and pressures reached values around \quant{3}{meV/\AA}  and \quant{1}{kBar}, respectively.  For the structural relaxations of the disordered systems, we used a 2$\times$2$\times$2 Monkhorst-Pack $k$-mesh with the 432 atom supercells in combination with Methfessel-Paxton\cite{Methfessel1989} Fermi surface smearing (\quant{\sigma=0.1}{eV}), while total energies and densities of states were calculated by the tetrahedron method with Bl\"ochl corrections\cite{Bloechl1994} using a 4$\times$4$\times$4 $k$-mesh. A 17$\times$17$\times$17 $k$-mesh was employed for the ordered structures represented in a cubic 16 atom unit cell. In all our calculations we allowed for a spontaneous spin polarization, always resulting in stable ferromagnetism.

In the SPR-KKR calculations, the ferromagnetic ground state was chosen as reference and disorder was treated analytically in the framework of the coherent potential approximation. The electronic density of states obtained for the different disordered structures agreed very well with the results obtained from the plane wave calculations, which corroborates our explicit supercell-based description.

\subsubsection{Formation energies and stable structures}
\begin{figure}[t]\centering
\includegraphics{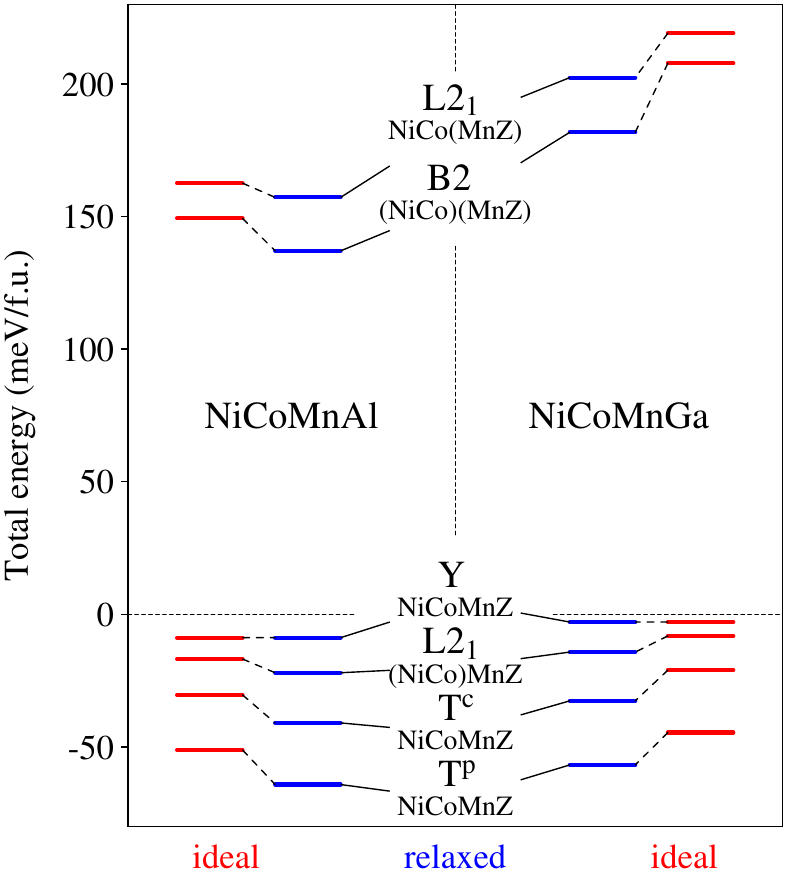}
\caption{Comparison of the total energies of the various structures of ferromagnetic NiCoMnAl and NiCoMnGa obtained from density functional theory. The inner levels (blue) refer to fully relaxed structures (positions and lattice constants), whereas in the outer columns (red) only the lattice constants were relaxed and the ions remain on the ideal (symmetric) positions with bcc coordination. The energies are specified relative to a mixture of the ferromagnetic ground state structures of ternary ferromagnetic Co$_2$MnZ and Ni$_2$MnZ. Thus negative energies denote structures which are inherently stable against demixing, whereas the others require to be stabilized by mixing entropy. \michael{idee: energie gegen volumen auftragen, relaxiert und unrelaxiert, Mn/Z ordnung durch volle pfeile, Ni/Co gestrichelt, relaxation gepunktet} \me{Würde das einen Verständnisgewinn bringen?} \michael{die zwei-dimensionale auftragung würde zeigen, dass das viereck der kubischen strukturen i.w. ein parallelogramm ist, dh. dass die ordnung des einen subgitters die gleiche volumensreduktion und energiegewinn bringt, egal ob das andere geordnet ist oder nicht. den energie-aspekt davon diskutiere ich ja schon im text. aber machen werd ich das erst einmal nicht} \michael{FIXME: abbildung}}
\label{fig:energy} 
\end{figure}

The results of our total-energy calculations are shown in Fig.\ \ref{fig:energy} and given in Tab.~\ref{tab:energies}. In addition to the quaternary systems, we also computed the ternary full Heusler systems for use as reference energies, specifically cubic L$2_1$  Co$_2$MnAl, Ni$_2$MnAl, and Co$_2$MnGa, as well as tetragonal L$1_0$ Ni$_2$MnGa, according to the martensitic transition occurring in the latter case. The energy differences are always specified with respect to the four-atom Heusler formula unit in the fully relaxed states. As expected for isoelectronic systems, the energy differences of the different phases behave similar in NiCoMnAl and NiCoMnGa. In both cases, we observe a significant gain in energy by ordering the main group element Z and Mn. As one would expect, the B2 phase is among the least favorable ones in terms of total energy, and thus its observed thermodynamic stability at high temperatures is due to its large configurational entropy. The fully disordered bcc phases turned out to be significantly higher in energy, at 1.00\,eV/f.u. for NiCoMnGa and 1.07\,eV/f.u. for NiCoMnAl (both without relaxation), and are therefore not included in Fig.\ \ref{fig:energy}. 

\begin{table*}[t]
\centering
\begin{tabular}{l|cc@{}c||@{\hspace{.5em}}c@{\hspace{1em}}c@{\hspace{1em}}c@{\hspace{.5em}}|rr@{\quad}|c|@{\hspace{.5em}}c@{\hspace{1em}}c@{\hspace{1em}}c}
& && & $a_0$ & $c_0$ &$V_0$ & $E_\text{form}$ &\multicolumn{1}{c|}{$\Delta E_\text{relax}$}& $M_\text{S}$ & $\mu_\text{Mn}$ & $\mu_\text{Ni}$ & $\mu_\text{Co}$ \\
    & \multicolumn{3}{c||@{\hspace{.5em}}}{Structure} &(\AA)&(\AA)&(\AA$^3$)&\multicolumn{2}{c|}{(meV/f.u.)}&($\mu_\text{B}$/f.u.)&\multicolumn{3}{c}{($\mu_\text{B}$/at.)}\\ \hline
NiCoMnAl&L2$_1$&NiCo  &(MnAl)&5.736&     &188.7&157&6&4.89&3.24&0.59&1.11\\
        &B2    &(NiCo)&(MnAl)&5.728&     &187.9&137&12&4.59&3.15&0.46&1.07\\
        &Y     &NiCo  &MnAl  &5.733&     &188.4&$-$9 &-&4.93&3.19&0.62&1.19\\
        &L2$_1$&(NiCo)&MnAl  &5.723&     &187.4&$-$22&5&4.60&3.10&0.51&1.10\\
        &\Tc   &      &      &5.662&5.839&187.2&$-$41&11&4.58&3.10&0.49&1.11\\
        &\Tp   &      &      &5.752&5.648&186.9&$-$64&13&4.43&3.05&0.40&1.11\\ \hline
NiCoMnGa&L2$_1$&NiCo  &(MnGa)&5.759&     &191.0&202&17&4.95&3.27&0.57&1.14\\
        &B2    &(NiCo)&(MnGa)&5.750&     &190.1&182&26&4.68&3.19&0.45&1.09\\
        &Y     &NiCo  &MnGa  &5.748&     &189.9& $-$3&-&4.97&3.23&0.60&1.18\\
        &L2$_1$&(NiCo)&MnGa  &5.738&     &188.9&$-$14&6&4.66&3.15&0.49&1.10\\
        &\Tc   &      &      &5.671&5.871&188.8&$-$33&12&4.61&3.14&0.46&1.09\\
        &\Tp   &      &      &5.772&5.658&188.5&$-$57&12&4.48&3.10&0.36&1.10
\end{tabular}
 \caption{Overview of the DFT results: Lattice constants and cell volumina, total energies of formation relative to the respective ternary Heusler compounds as discussed in the text, energy gain due to relaxation (resulting from disorder as well as tetragonal distortion), magnetic moments per formula unit NiCoMnZ and element-resolved moments for the various relaxed structures. For a better comparison, the lattice parameters $a_0$ and $c_0$ as well as $V_0$ refer in all cases to the 16 atom cubic cell instead of the respective primitive cell.}
\label{tab:energies}
\end{table*}

In contrast, the fully ordered Y structure, which has previously been proposed as a new candidate for a half metal, appears significantly more stable, not only against B2 disorder but also against decomposition into the ternary phases. However, a surprising result of our calculations is that NaCl-type ordering of Ni and Co is always disfavored compared to random disorder: this pertains both to L2$_1$ NiCo(MnZ), which is about \quant{20}{meV} higher in energy than B2 (NiCo)(MnZ), as well as the energetical gain of about \quant{12}{meV} when Y NiCoMnZ is disordered to L2$_1$ (NiCo)MnZ. Thus, considering only structural thermodynamics, the Y structure will not be thermodynamically stable at any temperature, as it has both higher internal energy as well as lower configurational entropy compared to L2$_1$ (NiCo)MnZ.

However, the partially disordered L2$_1$ structure should not be the ground state. Indeed, in both NiCoMnAl and NiCoMnGa the two tetragonal structures with a four-atom unit cell \Tp and \Tc have lower energies than all structures considered up to now. Thus, our calculations identify \Tp with the alternation of Ni and Co $(1,0,0)$ planes as the ground state structure. We are confident that, at least among the superstructures on the bcc lattice, there should be no structures with significantly lower energies, as the NaCl-type Mn/Z order with its large energy gain seems quite stable, while any Ni/Co order different from the three kinds considered here would need to rely on quite long-range interactions.

We observe that the relaxation procedure yields a considerable energy gain for the disordered structures. An analysis of the corresponding atomic displacements is given in the supplementary information, evidencing an expansion of $\langle 1,0,0\rangle$-coordinated pairs made up of equal atoms due to Pauli repulsion as common characteristic of the relaxations. Specifically for Mn/Z disorder, the mean bond lengths show an asymmetry, reflecting the larger size of the Z atom, particularly in the case for Z $=$ Ga. The relaxation energies of the disordered structures as given in Tab.~\ref{tab:energies} can be satisfactorily reproduced by assuming independent contributions of \quant{6}{meV} due Ni/Co disorder, \quant{6}{meV} due Mn/Al disorder, and \quant{18}{meV} due Mn/Ga disorder, with the prominence of the latter value again due to the larger size of Ga.

Due to the tetragonal arrangement of the Ni and Co atoms in \Tp and \Tc, the cubic symmetry is reduced to tetragonal, which is reflected also in the lattice parameters. Specifically, as reported in Tab.~\ref{tab:energies}, $c$, the lattice constant along the fourfold tetragonal axis, is 3--4\% larger than $a$ for the \Tc structure, while it is about 2\% smaller for \Tp. Indeed, this behavior is expected due to above-reported tendency of $\langle 1,0,0\rangle$-coordinated equal elements in L2$_1$ (NiCo)MnZ to be pushed apart due to Pauli repulsion, while Ni-Co pairs are contracted. Further, the Wyckoff positions 2c in space group 129, which are occupied by Mn and Z in the \Tp structure, have an internal degree of freedom $z$ correspond to a translation along the tetragonal axis. For NiCoMnAl, the parameters are $z_\text{Mn}=-0.2556$ and $z_\text{Al}=0.2512$, and for NiCoMnGa $z_\text{Mn}=-0.2555$ and $z_\text{Ga}=0.2511$, being practically the same in both compounds. With Ni in 2a at $z=0$ and Co in 2b at $z=1/2$, this means that Mn and Z are slightly shifted away from the Ni planes. Again, this is mirrored in the increased bond lengths of $\langle \tfrac{1}{2},\tfrac{1}{2},\tfrac{1}{2}\rangle$-coordinated Ni-Mn and Ni-Z pairs compared to Co-Mn and Co-Z pairs under disorder as given in the supplementary information. Thus, with these small tetragonal distortions and deviations of the internal degrees of freedom from the ideal values, it is clearly appropriate to consider also the tetragonal phases as superstructures on the bcc lattice.

While the tetragonal distortions as mentioned above are on the order of a few percent, the differences in the unit cell volumes between the cubic and the tetragonal structures is much smaller. Indeed, we observe that there is a nearly perfect monotonic decrease of unit cell volume with internal energy of the structures: while the volume contraction with Mn/Z ordering by values of about 0.2\% for NiCoMnAl and 0.6\% for NiCoMnGa was expected, and also the bigger effect in the latter case can be rationalized by the larger Ga atoms, NaCl-type Ni/Co order, which was already found to be energetically unfavorable, leads to a lattice expansion by about 0.5\% in both systems. In contrast, the energy gains with \Tc and \Tp are reflected in a corresponding volume contraction.

Our theoretical results explain the experimental observations: as reported above, experimentally NiCoMnGa displays the B2 phase at high temperatures with a well-defined ordering transition to the L2$_1$ (NiCo)MnGa phase at lower temperatures, while for NiCoMnAl the transition temperature is reduced and only barely kinetically accessible. This is reproduced by our calculations, with L2$_1$ (NiCo)MnZ being the lowest-energy cubic phase, while B2 can be stabilized by entropy, with indeed a larger energy gain and thus a higher expected transition temperature for Z = Ga. Ni/Co ordering always increases the internal energy and decreases configurational entropy, thus L2$_1$ NiCo(MnZ) and Y NiCoMnZ are not predicted to be existing. Also, as the energy cost of disordering B2 to A2 is about five times larger than the gain of ordering to L2$_1$ (NiCo)MnZ, with the latter happening around \quant{1000}{K}, we do not expect a transition to A2 in the stability range of the solid.

On the other hand, also the L2$_1$ (NiCo)MnZ phase is only stabilized by entropy, and thus should transform at some temperature to \Tp. With an argumentation as above, where the B2--L2$_1$ and the L2$_1$--\Tp transitions have the same entropy difference, but the latter's energy difference of \quant{42}{meV} is about a factor of 4--5 lower, we predict a transition temperature around \quant{250}{K} (see the supplementary material for a more detailed discussion of these issues). As already the B2--L2$_1$ transition in NiCoMnAl at around \quant{850}{K} is only barely progressing, a bulk transition to \Tp is therefore not to be expected on accessible timescales. 

Extrapolating the lattice constants measured on cooling to \quant{T=0}{K} gives \quant{5.777}{\AA} for NiCoMnGa and \quant{5.766}{\AA} for NiCoMnAl. The deviation of about \quant{0.04}{\AA} to the calculated values for L2$_1$ (NiCo)MnZ is quite satisfactory, corresponding to a relative error of 0.7\%. Of course, the difference in lattice constants between the two alloys should be predicted even much more accurately, giving \quant{0.015}{\AA} to be compared to the value of \quant{0.011}{\AA} determined experimentally. Further, the predicted contraction at the B2--L2$_1$ transition in NiCoMnGa of \quant{0.012}{\AA} agrees perfectly with the experimental value as obtained by integrating the excess thermal expansion coefficient between $1000$ and \quant{1200}{K}, while in NiCoMnAl the contraction with ordering is estimated as \quant{0.003}{\AA} by the differences in the heating and cooling curves evaluated at \quant{600}{K} and \quant{900}{K} to be compared with the predicted value of \quant{0.005}{\AA}. These two small discrepancies imply that the experimental lattice constant of NiCoMnAl at low temperatures on cooling is increased compared to the theoretical predictions, which is consistent with a reduced degree of L2$_1$ long-range order in NiCoMnAl due to kinetical reasons.
 
\subsubsection{Magnetism}
From the non-integer values of the total magnetic moment per formula unit listed in Tab.\ \ref{tab:energies}, one can already conclude that neither of the structures yields the desired half-metallic properties. For the fully ordered Y structure, our calculated values for $M_\text{S}$ are slightly lower than the values of \quantmub{5.0}\cite{Halder2015} and \quantmub{5.07}\cite{Alijani2011} previously reported for NiCoMnAl and NiCoMnGa, and the integer moment of 5\,$\mu_{\rm B}$, which follows from the generalized Slater-Pauling rule  $M_\text{S}=Z_\text{v}-24$ for half-metallic full-Heusler compounds with $Z_\text{v}=29$ valence electrons per unit cell. The calculated $M_\text{S}$ is even smaller for the other structures. Experimentally, we measured values between 4.47 and \quantmub{4.58} for L2$_1$ (NiCo)MnGa, while for Z $=$ Al an increase from \quantmub{4.68} in the as quenched state to \quantmub{4.89} after the longest annealing was observed. Thus, it seems that the dependence of $M_\text{S}$ on the state of order in the intermediate states is more complicated than the situation captured by our calculation of the respective extremes, corresponding to a decrease from perfect Mn/Z disorder in the B2 case to perfect order in the L2$_1$ case.

The values in Tab.\ \ref{tab:energies} imply that the magnetic moment per formula unit $M_\text{S}$ depends primarily on the order on the Ni/Co sublattice, with values around \quantmub{4.95} for NaCl-type order, \quantmub{4.60} for columnar order or disorder, and \quantmub{4.45} for planar order. The equilibrium unit cell volume $V_0$ grows with increasing $M_\text{S}$, with an additional lattice expansion in the cases of Mn/Z disorder. 

The induced Ni moments show the largest variation between the different structures, in absolute and relative numbers. The Co moments follow the behavior of the Ni moments with a smaller variation. This is a consequence of the hybridization of Co and Ni in the minority spin density of states, which is responsible for the formation of a gap-like feature at $E_\text{F}$ as discussed in detail in the next section. The Mn moments appear well localized with values slightly above \quantmub{3.1} and vary only by a tenth of a Bohr magneton.

The ferromagnetic ground state of the compounds arises from the strong ferromagnetic coupling between nearest-neighbor Mn-Ni (coupling constant approximately \quant{7}{meV}) and, in particular, Mn-Co (coupling constant approximately \quant{11.7}{meV}) pairs. On the other hand, Mn pairs in $\langle 1,0,0\rangle$ coordination, which randomly occur in the B2 case, exhibit large frustrated antiferromagnetic coupling (coupling constant approximately \quant{-8}{meV}). This behavior is well known from ternary stoichiometric and off-stoichiometric Mn-based Heusler systems.\cite{Sasioglu2004,Kurtulus2005,Sasioglu2005,Rusz2006,Buchelnikov2008,Sasioglu2008,Sokolovskiy2012,Comtesse2014,Entel2014} 
A more detailed account of the coupling constants for NiCoMnZ is given in the supplementary material. Thus for the low-energy structures, which do not exhibit Mn pairs with negative coupling constant, we expect the magnetic ordering temperature $T_\text{C}$ to be significantly higher than in the B2 case. This agrees nicely with the significant increase under annealing observed for NiCoMnAl, while NiCoMnGa is already L2$_1$-ordered in the as quenched state and thus has still a higher transition temperature.

\subsubsection{Electronic Structure}\label{sec:ElStruct}

The shape of the electronic density of states (DOS) of ternary L2$_1$ Heusler compounds of the type X$_2$YZ, including the appearance of a half-metallic gap, has been explained convincingly by Galanakis et al.\cite{Galanakis2002,Galanakis2006} in terms of a molecular orbital picture. First, we consider the formation of molecular orbitals on the simple cubic sublattice occupied by atoms of type X. Here, the $d_{xy}$, $d_{yz}$ and $d_{zx}$ orbitals hybridize forming a pair of \tgg and \tu molecular orbitals, while the $d_{x^2-y^2}$ and the $d_{3z^2-r^2}$ states form \eg and \eu molecular orbitals. The molecular orbitals of \tgg and \eg symmetry can hybridize with the respective orbitals of the nearest neighbor on the Y-position (in the present case Mn), splitting up in pairs of bonding and anti-bonding hybrid orbitals. However, due to their symmetry, no partner for hybridization is available for the \tu and \eu orbitals, which therefore remain sharp. Accordingly, these orbitals are dubbed ``non-bonding''.

If the band filling is adjusted such that the Fermi level is located between the \tu and \eu states in one spin channel, the compound can become half-metallic. This is for instance the case for Co$_2$MnGe with $Z_\text{v}=29$, which has according to the generalized Slater--Pauling rule $M_\text{S}=5\,\mu_\text{B}$.\cite{Picozzi2002,Galanakis2002} If additional valence electrons are made available, also the \eu states may become occupied. This is the case for Ni$_2$MnGa and  Ni$_2$MnAl ($Z_\text{v}=30$), which do not possess half-metallic properties. Here, the Ni-\eu states form a sharp peak just below the Fermi energy and gives rise to a band-Jahn-Teller mechanism leading to a martensitic transformation and modulated phases arising from strong electron-phonon coupling due to nesting features of the Fermi surface \cite{Brown1999,Lee2002,Bungaro2003,Zayak2003,Opeil2008,Haynes2012}. Consequently, the magnetic moments of these compounds are significantly smaller. First principles calculations report values of 3.97 -- 4.22\quantmub{} \cite{Ayuela1999,Godlevsky2001,Galanakis2002} and 4.02 -- 4.22\quantmub{}\cite{Fujii1989,Ayuela1999,Godlevsky2001,Ayuela2002,Sasioglu2004,Gruner2008} for Ni$_2$MnAl and Ni$_2$MnGa, respectively.

\begin{figure}[tb]
\includegraphics{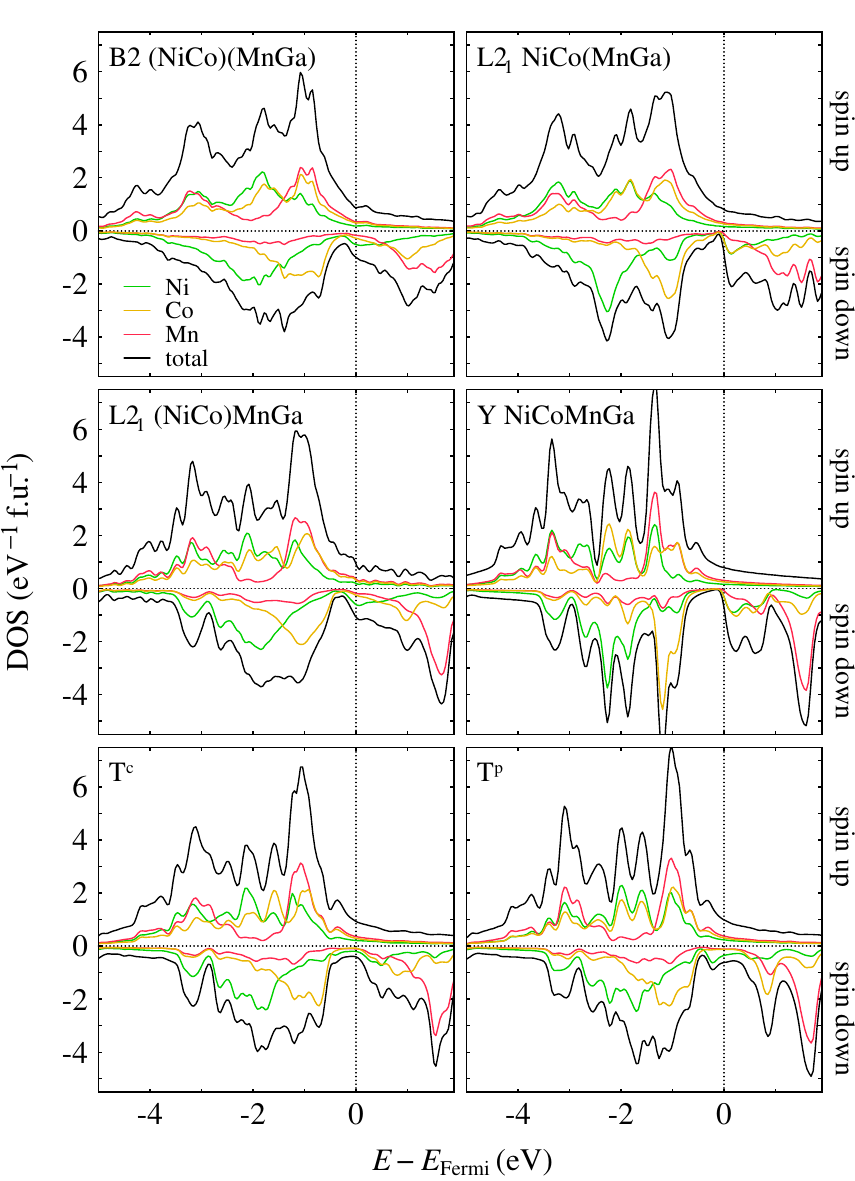}
\caption{Spin-polarized electronic density of states (DOS) in ferromagnetic NiCoMnGa for the distinct ordered structures.}
\label{fig:DOS} 
\end{figure}

Figure \ref{fig:DOS} shows the total and element-resolved electronic densities of states (DOS) of NiCoMnGa for the most relevant structures, which have the same valence electron concentration as the half metal Co$_2$MnGe (NiCoMnAl shows an analogous picture and can be found in the supporting material). Here, the perfectly alternating NaCl-type order of the elements on the Ni-Co sublattice in the Y-structure enforces a complete hybridization of the Ni- and Co-states, since the atoms find only $\langle 1,0,0\rangle$ neighbors of the other species. This becomes apparent from the pertinent illustration, as essentially the same features are present in the partial density of states of both elements. The magnitude of a specific peak may, however, be larger for one or the other species. This can be understood from the concept of covalent magnetism\cite{Williams1981,Schwarz1984,Mohn2006} which has been applied to Heusler alloys recently.\cite{Dannenberg2010} The molecular orbitals are occupied by each species with a weight scaling inversely with the energy difference to the constituting atomic levels. In the minority spin channel, the bonding molecular orbitals are dominated by Ni-states, while the non-bonding \tu states around \quant{-1}{eV} and the anti-bonding orbitals above $E_\text{F}$ are dominated by the Co-states. The non-bonding \eu orbitals directly above $E_\text{F}$ are equally shared by Co and Ni states. 

As expected, with decreasing order the features of the DOS smear out and become less sharp. Specifically the pseudo-gap at the Fermi level in one spin channel, which corresponds to the near half-metallic behavior suggested first by Entel et al.\ \cite{Entel2010}, and subsequently by Alijani et al.\ \cite{Alijani2011}, Singh et al.\ \cite{Singh2012} and Halder et al.\ \cite{Halder2015}, is in particular sensitive to ordering on the Ni-Co sublattice, and only encountered for the NaCl-type ordering of the fully ordered Y and the partially ordered L2$_1$ NiCo(MnZ) structures.

In fact, the minority spin gap is not complete. A close inspection of the band structure (see supporting material) of Y NiCoMnGa/Al clearly shows several bands crossing the Fermi level. Since this occurs in the immediate vicinity of the $\Gamma$ point, the weight of the respective states in the Brillouin zone is small and a gap-like feature appears in the DOS. Thus, in this configuration the compound should be classified as a half-semimetal rather than a half-metal. Nearly perfect gaps are observed if Z is a group IV element with a half-filled $sp$-shell. In our case, the missing electron of the main group element has to be compensated by the additional valence electron from one of the transition metals. These are only available on parts of the X-sites, which can lead to a distribution of $sp$-states between the sharp \tu and \eu states of the transition metals below and above $E_\text{F}$.

In all other structures, the Ni-Co sublattice contains neighboring pairs of the same element. In this case, the respective $d$-orbitals can hybridize independently at different levels. As a consequence, the \tu and \eu molecular orbitals split up. This is best seen in the DOS of the \Tp structure. Here, the Ni-dominated part of the former \eu peak moves to below the Fermi level (where we expect it in Ni$_2$MnGa/Al), which creates considerable DOS right at $E_\text{F}$ and fully destroys the half-metallic character. An analogous argument can be applied to the cubic L2$_1$ (NiCo)MnZ and B2 structures with disorder on the Ni-Co sublattice (Fig.\ \ref{fig:DOS}c and d). The disorder on the Mn-Z sublattice in the B2 phase causes only minor changes in the electronic structure and manifests mainly in a larger band width of the valence states and the disappearance of a pronounced peak at \quant{-3.2}{eV} in the minority channel, which originates from the hybridization between the Ni-Co and Mn-Z sublattice. In contrast, the distribution of the Ni and Co states near the Fermi level, which are decisive for the functional properties of this compound family, is not significantly changed compared to the L2$_1$ case.

\section{Conclusions}
Employing in-situ neutron diffraction, magnetic measurements and calorimetry, we studied the ordering tendencies in the quaternary Heusler derivatives NiCoMnAl and NiCoMnGa. NiCoMnGa was found to display an L2$_1$ (NiCo)MnGa structure with strong Mn/Ga order and no to minor Ni/Co ordering tendencies, where the degree of order achieved upon slow cooling was higher than in quenched samples. The B2--L2$_1$ second-order phase transition was observed at \quant{1160}{K}. NiCoMnAl after quenching was found to adopt the B2 structure, while on slow cooling from high temperatures broadened L2$_1$ reflections were observed to emerge at temperatures below \quant{850}{K} in neutron diffractometry. Yet, kinetics at these temperatures are so slow that the adjustment of large degrees of L2$_1$ order in this compound is kinetically hindered. Still, low-temperature annealing treatments at \quant{623}{K} in samples quenched from \quant{1273}{K} showed a strong effect on the magnetic transition temperatures, proving that this parameter probes sensitively the state of order in the sample.

Density functional theory reproduces the experimentally observed trends of the order-dependent magnetic behavior and of the ordering tendencies between the two systems. Our calculations reveal that the fully ordered Y structure with F$\overline{4}$3m symmetry is thermodynamically not accessible, since the partially disordered L2$_1$ phase is lower in energy. Instead, we propose as the ground state a tetragonal structure with a planar arrangement of Ni and Co. This structure is stable against decomposition into the ternary Heusler compounds, but we expect the energetic advantage to be too small to compensate for the larger entropy of the L2$_1$ phase at reasonable annealing conditions. However, the fabrication of this structure by layered epitaxial growth on appropriately matching substrates, which favor the slight tetragonal distortion, could be possible. From the electronic density of states and band structure, we could conclude that neither of the structures is half-metallic in the strict definition. This specifically pertains also to the hypothetical Y structure, which exhibits several bands crossing the Fermi level close to the $\Gamma$ point in the minority spin channel, and is thus a half-semimetal.

Since the first quaternary Heusler derivatives adopting the Y structure have been proposed to possess half-metallic properties\cite{Dai2009}, the interest in these materials has developed rapidly with numerous publications dealing with the topic.\cite{Entel2010,Alijani2011, Goekoglu2012,Singh2012,Alzyadi2015,Wei2015,Halder2015,
Mukadam2016,
Alijani2011b,Gao2013,Ozdougan2013,Zhang2014,Bainsla2014,Xiong2014,Enamullah2015,Feng2015,Gao2015,Elahmar2015, Enamullah2016,
Berri2014} Density-functional theory calculations have been used to identify promising systems among the NiCo-,\cite{Entel2010,Alijani2011, Goekoglu2012,Singh2012,Alzyadi2015,Wei2015,Halder2015} NiFe-\cite{Alijani2011,Wei2015,Mukadam2016} and CoFe-based\cite{Dai2009,Alijani2011b,Gao2013,Ozdougan2013,Berri2014,Zhang2014,Bainsla2014,Xiong2014,Enamullah2015, Feng2015,Gao2015,Elahmar2015,Enamullah2016} compounds. However, in most cases the phase stability of the Y structure is tested, if at all, only against stacking order variations of this Y structure (see, for instance, Refs.\ \onlinecite{Alijani2011b,Dai2009}) but rarely against disorder\cite{Enamullah2016} or other states of order. Simultaneously, experimental investigations as a rule either point towards disordered structures \cite{Alijani2011b,Bainsla2014,Enamullah2015, Halder2015,Mukadam2016} or, specifically for the case of X-ray diffraction on ordering between transition metal elements, cannot decide these issues.\cite{Alijani2011,Alijani2011b,Enamullah2015}

Based on our findings, we conclude that at least in the NiCo-based, but probably also in the NiFe- and CoFe-based alloys, the stability of the Y-structure is doubtful and, even if it was thermodynamically stable, might still not be kinetically accessible in most quaternary Heusler derivatives. Indeed, preliminary first-principles results show that also for the NiFeMnGa and the CoFeMnGa alloys, the tetragonal \Tp order is lower in energy than the Y structure by 62\,meV/f.u and 80\,meV/f.u., respectively. This underlines that a detailed analysis of phase stabilities in those systems that have been identified as promising half-metals, especially with respect to the tetragonal structures and/or L2$_1$ type disorder, is essential in order to evaluate their actual potential. More generally, the comparatively small energetical differences between the various possible types of order along with small disordering energies specifically with respect to the late transition metal constituents as obtained here suggest that in these quaternary Heusler derivatives disorder could be the norm rather than the exception in physical reality.

\section{Acknowledgments}
This work was funded by the Deutsche Forschungsgemeinschaft (DFG) within the Transregional Collaborative Research Center TRR 80 ``From electronic correlations to functionality''. P.N.{} acknowledges additional support from the Japanese Society for the Promotion of Science (JSPS) via a short-term doctoral scholarship for research in Japan. We thank O.{} Dolotko and A.{} Senyshyn of the MLZ for facilitating the neutron diffraction measurements. SQUID measurements were performed at the Center for Low Temperature Science, Institute for Materials Research, Tohoku University. Computing resources for the supercell calculations were kindly provided by the Center for Computational Sciences and Simulation (CCSS) at University of Duisburg-Essen on the supercomputer magnitUDE (DFG grants INST 20876/209-1 FUGG and INST 20876/243-1 FUGG).

\end{document}